\documentclass[twocolumn,showpacs,preprintnumbers,amsmath,amssymb]{revtex4}
\usepackage{graphicx}
\usepackage{dcolumn}
\usepackage{bm}
\input epsf
                                                               
\begin{document}
\preprint{}
\title{Long range scattering resonances in strong-field seeking
states of polar molecules}
\author{Christopher Ticknor and John L. Bohn}
\affiliation{JILA, National Institute of Standards and Technology 
and University of Colorado, Boulder, Colorado 80309-0440, USA
}
\date{\today}
                                                                         
\begin{abstract}
We present first steps toward understanding the ultracold scattering properties
of polar molecules in strong electric field-seeking states.  We have found
that the elastic cross section displays a quasi-regular set of potential 
resonances as a function of the electric field, which potentially
offers intimate details about the inter-molecular interaction.
We illustrate these resonances in a ``toy'' model composed of pure dipoles,
and in more physically realistic systems.  To analyze these resonances, we use 
a simple WKB approximation to the eigenphase, which proves
both reasonably accurate and meaningful.  A general treatment of 
the Stark effect and dipolar interactions is also presented.
\end{abstract}
\maketitle
\section{Introduction}
Conventional spectroscopy of atoms and molecules begins with
the premise that the energy levels of the species being probed are 
fixed although they may shift slightly in electromagnetic fields.  
These levels are then interrogated by energy-dependent
probes such as photons or charged or neutral particles.  Information 
on the energetics and structure of the molecules is extracted from 
absorption energies, oscillator strengths, selection rules, etc.  
In these investigations, the study of resonances has 
played a central role.

With the advent of ultracold environments for atoms and molecules, 
this general view of spectroscopy can be inverted.  Cold collisions 
provide a nearly monochromatic probe of near-threshold intermolecular 
interactions, with resolution set by the milliKelvin or microKelvin 
temperature of the gas.  In this case the energy levels of nearby 
resonant states can be tuned into resonance with the zero-energy 
collisions.  For cold gases of alkali atoms, this strategy is already in 
widespread use.  It exploits the fact that the Zeeman effect can shift 
the internal energies of the atoms over ranges orders of magnitude 
larger than the collision energy itself.  In this way, atoms can be 
made to resonate when they would not naturally do so (i.e., in zero field).  
Measurement of these ``Feshbach'' resonances (more properly, Fano-Feshbach 
resonances) has yielded the most accurate determination of 
alkali-alkali potential energy surfaces for near-threshold
processes \cite{potentials}.

The current experimental and theoretical push to create and study 
ultracold molecules \cite{cold-rev} will lead to many more 
opportunities for this novel kind of spectroscopy, since molecules 
possess many more internal degrees of freedom than do atoms.  There 
will be, for example, numerous Fano-Feshbach resonances in which one or 
both of 
the collision partners becomes vibrationally or rotationally excited 
\cite{Forrey,Deskevich,Bala}; statistical arguments suggest that 
these resonances will be quite narrow in energy, a fact related to 
their abundance \cite{Deskevich}.
A second class of resonant states will occur when the constituents 
are excited into higher-lying fine structure or hyperfine structure 
states, more reminiscent of the resonances observed in the 
alkali atoms.  In molecules, these resonances are naturally also tunable 
in energy using magnetic fields \cite{krems}.

In this paper, we are primarily interested in a third class of resonance:
potential resonances that are engendered by altering the 
intermolecular potential energy surface itself.
This capability becomes especially prominent in cold collisions of 
heteronuclear polar molecules, whose dipolar interactions are quite 
strong on the scale of the low translational temperatures
of the gas.  At the same time, the dipole moments of the individual 
dipoles can be strengthened or weakened as well as aligned
by an applied electric field.

This kind of resonance includes the shape resonances, discussed 
in the context of cold atoms polarized by 
strong electric fields \cite{shape-you,Deb}.  
A second set, dubbed ``field-linked'' resonances, has been studied in 
some detail in Refs \cite{AA_PRL,AA_PRA,AA_FL}.  These states appear in 
potential energy surfaces (PES's) that correlate 
to weak electric field-seeking states of the free molecules.  They are 
weakly bound, long-ranged in character, and indeed do not appear to 
exist without an electric field present.  Similar resonances 
are predicted to occur in 
metastable states of the alkaline earth elements at low temperatures 
\cite{derev,santra}.

A rich set of potential resonances emerges among strong-field
seeking states, and this is the subject of the present paper.  Strong-field 
seekers are of increasing importance experimentally, since they would 
enable molecules to be trapped
in their absolute ground states where no 2-body inelastic collision processes 
are available to harm the gas.  In this case, the colliding 
molecules are free to approach to within a small internuclear distance of one 
another; the resulting potential resonances therefore can probe detailed 
intermolecular dynamics near threshold.  The resulting data, consisting
of scattering peaks as a function of electric field, can be thought of
as a kind of ``Stark spectroscopy''.
Just such a tool has been applied previously in precision measurements of
alkali Rydberg spectra \cite{Ryd1,Ryd2}.

In this paper we explore such spectra in ultracold molecules, finding  that
the spectrum is dominated by a quasi-regular series in the electric
field values.  Such a series is the fundamental building block of molecular
Stark spectroscopy and plays a role analogous to the Rydberg
series in atomic spectroscopy.  In both cases, the series
lays out the fundamental structure of the unperturbed, long-range
physics between interacting entities.  In the case of atoms,
the {\it deviation} from an unperturbed Rydberg series, encapsulated in the
quantum defects, yields information on the electron-core interaction 
\cite{Seaton,qdt}. Similarly, it is expected that differences in observed 
Stark 
spectra from those presented here will probe the short-range intermolecular
interactions.

To emphasize the generality of these resonances, we first briefly develop 
in Sec. II a general formalism for arbitrary polar collision partners, 
covering explicitly Hund's cases (a) and (b) as well as asymmetric rotor
molecules.  Using this formalism, we introduce in Sec. III the 
potential resonances using a simplified version of the molecular gas in 
which all dipoles are assumed to be perfectly aligned and where molecular fine 
structure plays no role.  In Sec. IV, we consider the case of more
realistic molecules, where fine structure does intervene.  We
show that the structure of the potential resonances is unchanged,
but that additional
narrow Fano-Feshbach resonances do appear.  Throughout the article we 
emphasize 
how the various types of resonance can be classified and organized by simple 
considerations involving the WKB approximation.

\section{General Form of the Stark Effect and dipolar interaction}
In this section, we briefly recapitulate our scattering model which is
presented in Refs.\cite{AA_PRA,CT1,CT_thesis}.  
The molecular scattering Hamiltonian 
can be written as
\begin{equation}
H=T_0+V_{SR}+V_{LR}+H_{S}+H_{int}\label{ham}
\end{equation}
Where $T_0$ is the kinetic energy operator, $V_{SR}$ is the short range
potential energy surface
(PES), $V_{LR}$ is the long range interaction of the polar molecules, 
$H_{S}$ is the Stark Hamiltonian, and $H_{int}$ is the Hamiltonian 
describing the internal degrees of freedom of the 
molecules. 
For the purpose of this paper we disregard $V_{SR}$.
The long range interaction is dominated by the dipole-dipole
interaction \cite{AA_PRA}.  Furthermore an electric field can significantly 
change the structure of the interaction and be used to control the molecular 
collisions \cite{AA_PRA,CT1}. 

The first step in performing a scattering calculation is to 
construct a basis of molecular energy eigenstates in an external field.
Here we generalize the approach slightly to make it applicable
to various types of polar molecules.  Once the Stark Hamiltonian is 
obtained, the result can be used in the Hamiltonian of the dipolar interaction,
showing their common physical origin in terms of electric fields.

To simplify the analysis and calculations, we assume that the vibrational 
degrees of freedom are frozen out at low temperatures; hence can treat
the molecules as rigid rotors.  The molecular state is described in terms 
of the state $|JM_J\Omega\rangle$, where $J$ is the molecule's rotational
plus electronic angular momenta, $M_J$ is the projection 
of $J$ onto the lab axis, and $\Omega$ is $J$'s projection onto the 
molecular axis.
To describe the rigid rotor molecular wave function, we use 
$\langle \alpha,\beta,\gamma |J M_J \Omega \rangle=\sqrt{2 J+1\over 8
\pi^2}D^{J\star}_{M_J\Omega}(\alpha,\beta,\gamma)$,
where $\alpha,\beta,$ and $\gamma$ are the Euler angles defining the molecular
axis and $D^{J\star}_{M_J\Omega}$ is a Wigner D-function \cite{Brown}.

\subsection{The Stark effect}
Since the electric field is a true vector (as opposed to
a pseudovector), it only couples states of opposite parity.
This implies that the Stark energies vary quadratically 
with low electric field; 
they vary linearly only at higher fields once the Stark 
energy is greater than the energy splitting of the two states.
The Stark Hamiltonian has the form
\begin{equation}
H_{S}=-{\vec\mu}\cdot \vec{\cal E}. 
\label{Stark}
\end{equation}
For the current discussion, we pursue a general approach and allow the 
field to point in any direction in the laboratory reference frame.  
This general approach allows the 
matrix elements to be used in the dipolar interaction.  However when 
explicitly considering the molecular states in an electric field, we 
that the field lies in the $\hat z$ direction.  

The Stark interaction 
can be evaluated by decomposing the electric field in
spherical coordinates and then rotating the dipole into the lab frame.
In this representation, the Stark Hamiltonian has the form $H_S=-\sum_q\mu_{} 
{\cal E} D^{1\star}_{q0}$, where $\mu_{}$ is the electric dipole moment of 
the molecule and $D^{1\star}$ is a Wigner D-function. $D^{1\star}_{q0}$ is 
equal to $(-1)^{2q}C^1_q$, a reduced spherical harmonic \cite{Brink}, 
and $q$ is the projection quantum number of $\cal E$ onto the 
lab axis.  At the heart of evaluating the Stark effect, we find the 
operator $D^{1\star}$ coupling two molecular states, which themselves
are described by D functions.  The Stark matrix element is an integral of 
three D-functions over the molecular coordinates, averaging over molecular 
orientation.  The evaluation of the integral 
results in selection rules whose details depend on the molecular specifics.  
For a valuable qualitative discussion on the Stark effect see \cite{Schreel}.

A given molecule may also have a nuclear spin that
generates a hyperfine structure.  In this case, it is more appropriate 
to present the matrix elements in the hyperfine basis, where
$F$ and $M_F$ define the state. Here $F$ is the sum of $J$ and the nuclear
spin $I$ and $M_F$ is the projection of $F$ in the lab frame.
We use the Wigner-Eckart theorem to compute the Stark 
matrix elements in a compact form. The matrix elements of the Stark effect 
are written as
\begin{eqnarray}
\langle \alpha F M_F |H_{S} |\alpha^\prime F^\prime M_F^\prime\rangle
=-\mu_{}{\cal E} \langle \alpha FM_F | D^{1\star}_{q0}| F^\prime M_F^\prime
\alpha^\prime \rangle,
\end{eqnarray}
which contains a purely geometrical matrix element
\begin{eqnarray}
\label{dipole_matrix_element}
\langle \alpha FM_F | D^{1\star}_{q0}| F^\prime M_F^\prime
\alpha^\prime \rangle = 
[F](-1)^{1+M_F+F^\prime} \nonumber\\ \times
\left(\begin{array}{ccc}
F^\prime&1&F\\
M_F^\prime&M_F-M_F^\prime&-M_F
\end{array}\right)
\langle \alpha F\parallel  D^{1\star}_0\parallel\alpha^\prime F^\prime \rangle.
\label{stark}
\end{eqnarray}
Here 
$\langle \alpha F\parallel D^{1\star}_0\parallel\alpha^\prime F\prime \rangle$
is the reduced matrix element and $\alpha$ represents all remaining 
quantum numbers needed to uniquely determine the quantum state.  
$[j]$ is a shorthand notation for $\sqrt{2j+1}$.  We have left $q$ in the
matrix element simply as a place holder.  Its value is assumed to be
$q=M_F-M_F^\prime$ in accordance with conservation of angular momentum.
If we consider the case where the electric field points
in the $\hat z$ direction, then $q=0$, and we find that $M_F=M_F^\prime$.
This means that $M_F$ is a conserved quantum number.   
We have used the convention of Brink
and Satchler to define the Wigner-Eckart theorem and reduced matrix elements
\cite{Brink}.

\subsection{Molecular examples}
We present a few examples of reduced matrix elements for specific molecular 
symmetries.  First, consider a Hund's case (a) molecule with $\Omega \ne 0$.  
The OH radical, with ground state $^2 \Pi_{3/2}$, is a good example of this. 
The energy eigenstates of
such a molecule in zero electric field are eigenstates of parity, i.e.,
$|J M_J\bar \Omega\epsilon \rangle$  where $\epsilon=+(-)$ represents 
the $e$ ($f$) parity state and $\bar\Omega=|\Omega|$. 
The parity of this molecule is $\epsilon (-1)^{J-1/2}$ if $J$ is a half 
integer or $\epsilon (-1)^{J}$ if $J$ is an integer.
(For details see Refs. \cite{Schreel,CT1,CT_thesis}.)
Taking  this molecular structure into account, we find that the 
reduced matrix element is
\begin{eqnarray}
\langle \alpha F \parallel D^{1\star}_0\parallel F^\prime\alpha^\prime\rangle= 
(-)^{1+I+F+J+J^\prime-\bar\Omega}[F^\prime,J,J^\prime]\nonumber\\\times
\left\{\begin{array}{ccc}F&F^\prime&1\\J^\prime&J&I\end{array}\right\}
\left(\begin{array}{ccc}J^\prime&1&J\\
-\bar\Omega&0&\bar\Omega\end{array}\right)\nonumber\\\times
\left({1+\epsilon\epsilon^\prime(-1)^{J+J^\prime+2\bar\Omega+1}\over 2}
\right)\label{redOH}.
\end{eqnarray}
Here the index $\alpha$ represents the set of quantum numbers
$\epsilon,\bar \Omega$ and $J$.
We have introduced another notation: $[j_1, j_2, ...,j_N] =$ 
$\sqrt{(2j_1+1)(2j_2+1)\cdots(2j_N+1)}$.  
This reduced matrix element is the same for any case (a) molecule.

On the other hand consider a Hund's case (b) molecule with $L = 0$.  Many 
molecules fit this mold such as heteronuclear alkali dimers and SrO
with $^{2S+1}\Sigma$ ground states.  Here the parity of a state
is directly identified by the value of $J$, where
$parity = (-1)^{J}$.
The Stark effect therefore mixes the ground state with the next 
rotational state.  We find the reduced matrix element to be 
\begin{eqnarray}
\langle \alpha F\parallel D^{1\star}_0\parallel F^\prime\alpha^\prime\rangle= 
(-1)^{F+I+J+J^\prime+S+N+N^\prime}\nonumber\\\times
{[N,N^\prime,J,J^\prime,F^\prime]}
\left\{\begin{array}{ccc}F&F^\prime&1\\J^\prime&J&I\end{array}\right\}
\nonumber\\\times
\left\{\begin{array}{ccc}J&J^\prime&1\\N^\prime&N&S\end{array}\right\}
\left(\begin{array}{ccc}N^\prime&1&N\\0&0&0\end{array}\right)
\label{redSRO}
\end{eqnarray}
where the index $\alpha$ represents  the set of quantum numbers
$N$ and $S$.  

This formalism is easily extended to include asymmetric rotors by
including the rotational Hamiltonian to construct molecular 
eigenstates.  For an asymmetric rotor, there are
three distinct moments of inertia and therefore three distinct rotational 
constants.  The rotational Hamiltonian is
$H_{rot}=A{\bf J}^2_a+B{\bf J}^2_b+C{\bf J}^2_c$, where $A>B>C$ and
$a,b$ and $c$ are the axis labels in the molecular frame. 
This additional structure mixes $\Omega$ such that it is not
a good quantum number, implying we that need to diagonalize 
the rotational Hamiltonian along with the Stark Hamiltonian to obtain the 
molecular eigenstates.
We consider the primary effect of this additional structure to change 
the progression of rotationally excited states.
For a complete discussion on asymmetric rotors see Refs. \cite{Hain,tinkham}.

\subsection{Dipole-dipole interaction}
The intrigue of polar molecules is their long range anisotropic scattering
properties whose origin is the dipole-dipole interaction.
The interaction in vector form is
\begin{equation}
 H_{\mu\mu}=-\
{3 (\hat {\bf R} \cdot  \hat{\bf \mu}_1)(\hat {\bf R} \cdot \hat
{\bf \mu}_2)-\hat{\bf \mu}_1 \cdot\hat{\bf\mu}_2 \over R^3},
\label{fulldidi}
\end{equation}
where ${\bf \hat \mu}_i$ is the electric dipole of molecule $i$, $R$ is the
intermolecular separation, and ${\bf\hat R}$ is the unit vector defining
the intermolecular axis.  This interaction is conveniently
rewritten in terms of tensorial operators in the laboratory frame as:
\begin{equation}
\label{didi}
 H_{\mu\mu}= -{\sqrt{6}\over R^3} \sum_{q} (-1)^q
C^2_{-q} \cdot (\mu_1 \otimes  \mu_2)^2_{q}.
\end{equation}
Here $C^2_{-q}(\theta,\phi)$ is a reduced spherical harmonic that
acts on the relative angular coordinate of the molecules,
while $(\mu_1 \otimes  \mu_2)^2_{q}$ is the second rank
tensor formed from the two rank one operators determining the individual
dipoles.  For this reason, matrix elements of
the interaction are also given conveniently in terms of the
matrix elements in Eq. (\ref{dipole_matrix_element}).  The matrix 
elements of the dipolar interaction are:
\begin{eqnarray}
\langle 1 2 l m_l|H_{\mu\mu}| 1^\prime 2^\prime l^\prime m_l^\prime\rangle
= (-1)^{-m_l+m_l^\prime+1}
\left({\mu_{}^2\sqrt{6}\over R^3}\right)\nonumber\\\times
\langle lm_l|C_{(m_l-m_l^\prime)}^2|l^\prime m_l^\prime\rangle
\langle \alpha_1 F_1 M_{F_1} | D^{1\star}_{q0}| \alpha_1^\prime 
F_1^\prime M_{F_1}^\prime \rangle
\nonumber\\\times
\langle \alpha_2 F_2 M_{F_2} | D^{1\star}_{q^\prime0}| \alpha_2^\prime 
F_2^\prime M_{F_2}^\prime \rangle 
\nonumber\\\times
\left(\begin{array}{ccc}1&1&2\\
M_{F_1}-M_{F_1}^\prime&M_{F_2}-M_{F_2}^\prime&m_l-m_l^\prime\end{array}\right)
\label{matdd}
\end{eqnarray}
where 
\begin{eqnarray}
\langle lm_l|C_{(m_l-m_l^\prime)}^2|l^\prime m_l^\prime\rangle=(-1)^{m_l}
[l,l^\prime]
\left(\begin{array}{ccc}l&2&l^\prime\\0&0&0\end{array}\right)\nonumber\\\times
\left(\begin{array}{ccc}l&2&l^\prime\\
-m_l&m_l-m_l^\prime&m_l^\prime\end{array}\right).
\end{eqnarray}
Equation (\ref{matdd}) shows that once the Stark Hamiltonian has been 
constructed
for a particular molecule, then the Hamiltonian describing the dipolar 
interaction can be achieved with little extra effort.  
This result reflects the fact that in Eq. (\ref{matdd}) each dipole is acted 
on by the electric field of the other.  We have also used a shorthand to 
represent the channel, $|12lm_l\rangle$,  where $|1\rangle$ denotes the 
quantum state of the first molecule and as  $|2\rangle$ for second 
molecule.  At this point, we disregard all interactions between molecules
except the dipolar interaction.

To exploit an analogy with Rydberg atoms, the long range Coulombic
interaction is well understood. With this understanding of the long range 
physics, a solution to the complete problem is achieved by matching it to 
the short range solution
or a parameterization of the short range interaction \cite{qdt}.  
This idea was, in fact, implemented as a numerical tool in Ref.\cite{Deb},
which dealt with collisions of atoms in strong electric fields.
To this end, we pursue an understanding of the long range characteristics of
the dipolar scattering.  We then envision merging the long range physics with
the short range physics, or a parametrization thereof, to offer insight into
the short range interaction and to explore the dynamic interaction of
polar molecules.  With this idea in mind we first explore  ``pure'' 
dipolar scattering.

\section{Dipolar Scattering}
\begin{figure}
\centerline{\epsfxsize=7.0cm\epsfysize=7.0cm\epsfbox{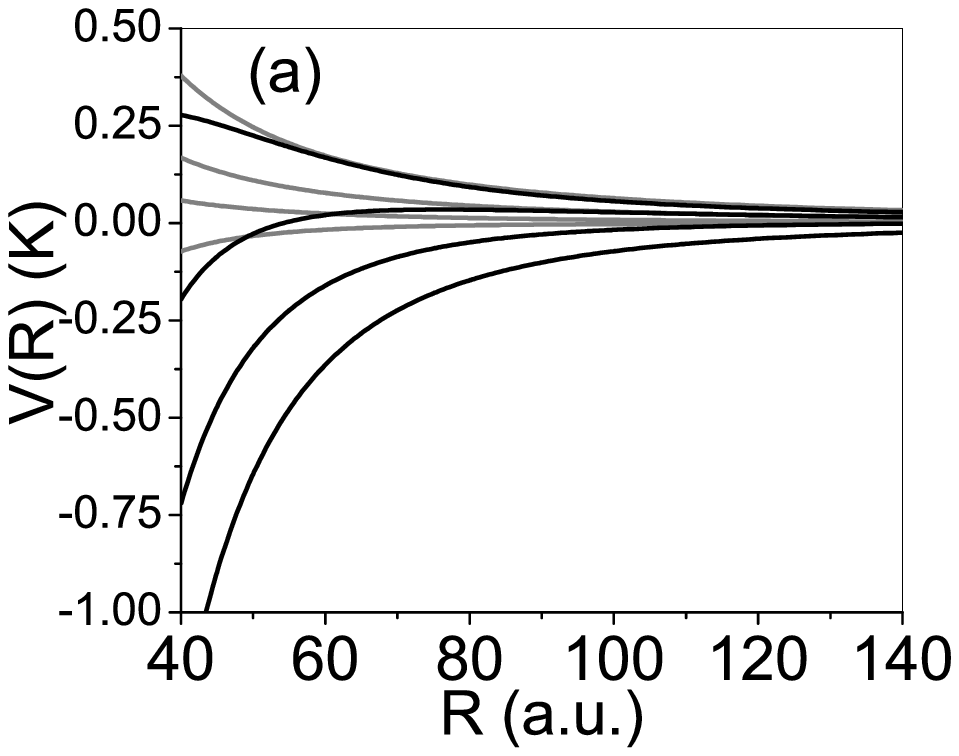}}
\centerline{\epsfxsize=7.0cm\epsfysize=7.0cm\epsfbox{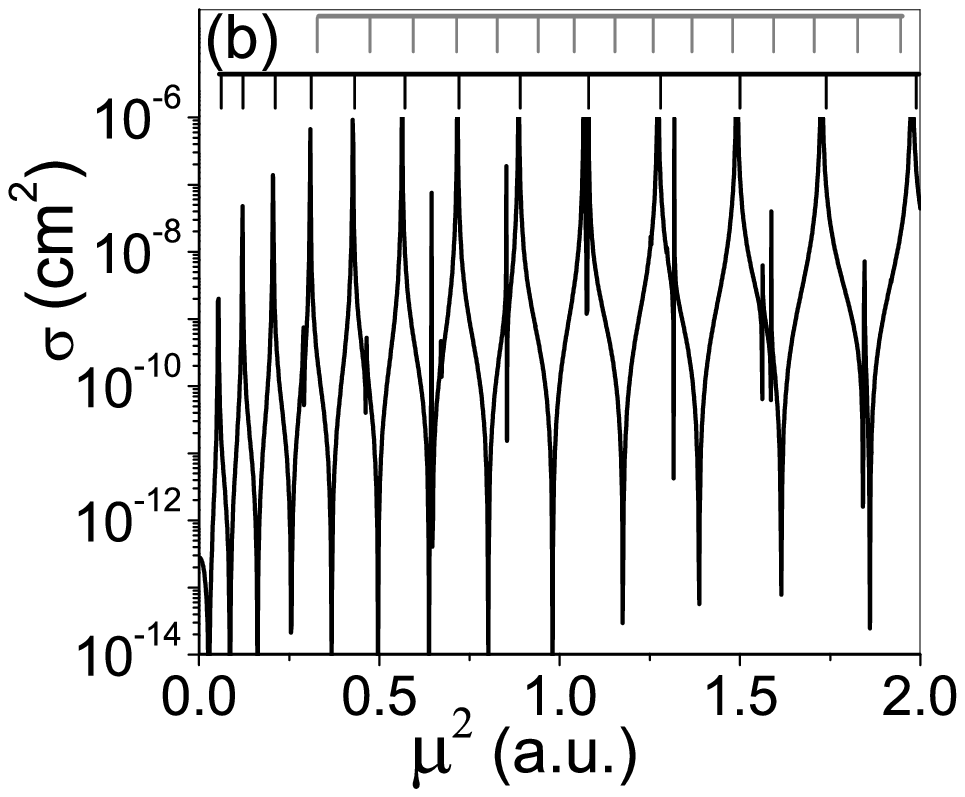}}
\caption{(a) Adiabatic curves of the pure dipole system for
two values of the dipole moment $\mu$: $\mu=0.1$ (gray) and $\mu=1$ (black) 
(b) The cross section of the polarized dipole model
versus $\mu$, at a collision energy $E=10^{-12}$ K.
The brackets denote predicted resonant positions using the 
adiabatic WKB phase (AWP) approximation.  The black bracket represents 
for the phase contribution from the lowest adiabatic curve and the gray is 
from all remaining contributions.}\label{model-sigma}\end{figure}

Our primary interest in polar molecule scattering is how the strong
anisotropic interaction affects the system.  As a first step 
illustrating the influence of dipolar interactions, we present
a simple model composed of polarized dipoles with no internal structure.
Strictly speaking, this system is created by an infinitely strong 
electric field that completely polarizes the molecules and 
raises all other internal states to experimentally unattainable energies.
Thus the only label required for a channel is its 
partial wave, $l=\{0,2,4,...\}$ in the numerical example given here. 
The matrix elements of the dipole-dipole 
interaction are taken to be $\langle  12l0|H_{\mu\mu}|12l^\prime0\rangle
=-0.32\mu^2\langle l0|C_0^2|l^\prime0\rangle$, which is typical for
molecules like RbCs or SrO.  We then artificially vary the dipole 
moment $\mu$ to see the effect of an increasingly 
strong dipolar interaction.  Pragmatically speaking, varying $\mu$ parallels 
changing the electric field.  The intention of this model is to focus on the 
effect of direct anisotropic couplings between the degenerate channels,
as measured by their effect on the partial wave channels.

For this model, we use a reduced mass of $m_r=10^4$ a.u., typical
of very light molecules.  We moreover assume that the molecules
approach one another along the laboratory z-axis, so that only the
projection $m_l=0$ of orbital angular momentum is relevant.
To set a concrete boundary condition at small $R$, we apply ``hard
wall'' boundary conditions, $\psi(R_{in})=0$ at a characteristic 
radius $R_{in}=20$ a.u.  (In Sec. V we will relax this restriction.)
We pick the collision energy to be nearly zero, namely $10^{-12}$ K.
To converge the calculation for this model requires inclusion of
partial waves up to $l=14$, and numerical integration of the
Schr\"{o}dinger equation out to $R=R_{\infty}=1\times10^{5}$ a.u.,
using the log-derivative propagator method of Johnson \cite{Johnson}.

To get a sense of the influence of increasing the dipole moment,
we first look at adiabatic curves of the system. 
Figure \ref{model-sigma} (a) shows two different sets of adiabatic curves:
a gray set with $\mu=0.1$ (a.u.) and a black set $\mu=1.0$ (a.u.).  In each 
set, the four lowest adiabatic curves are shown.  Looking at these curves, 
we can see two characteristic effects of increasing $\mu$.
First, the lowest curve becomes much deeper.
Second, the higher adiabatic curves, originating form non-zero 
partial waves, may support bound states at short distance states, i.e. 
within the centrifugal barrier.  Both these effect may generate bound states,
leading to distinct classes of scattering resonances as $\mu$ is varied.  
The deepening of the lowest adiabatic curves induces potential 
resonances, whereas the higher-lying curves lead to narrow shape resonances,
wherein the molecules must tunnel through the centrifugal barrier.

The different classes of resonances can clearly be seen in cross section,
as shown in Fig. \ref{model-sigma} (b).
The broad quasi-regular set of resonances seen in the cross section 
are the potential resonances originating from the lowest adiabatic curve.
The narrow shape resonances appearing sporadically in the spectrum originate 
from the higher-lying curves.  For the purpose of this paper,
we focus on the wide potential resonances and simply acknowledge
the existence of the narrow shape resonances. 

\begin{figure}\centerline{\epsfxsize=7.0cm\epsfysize=7.0cm\epsfbox{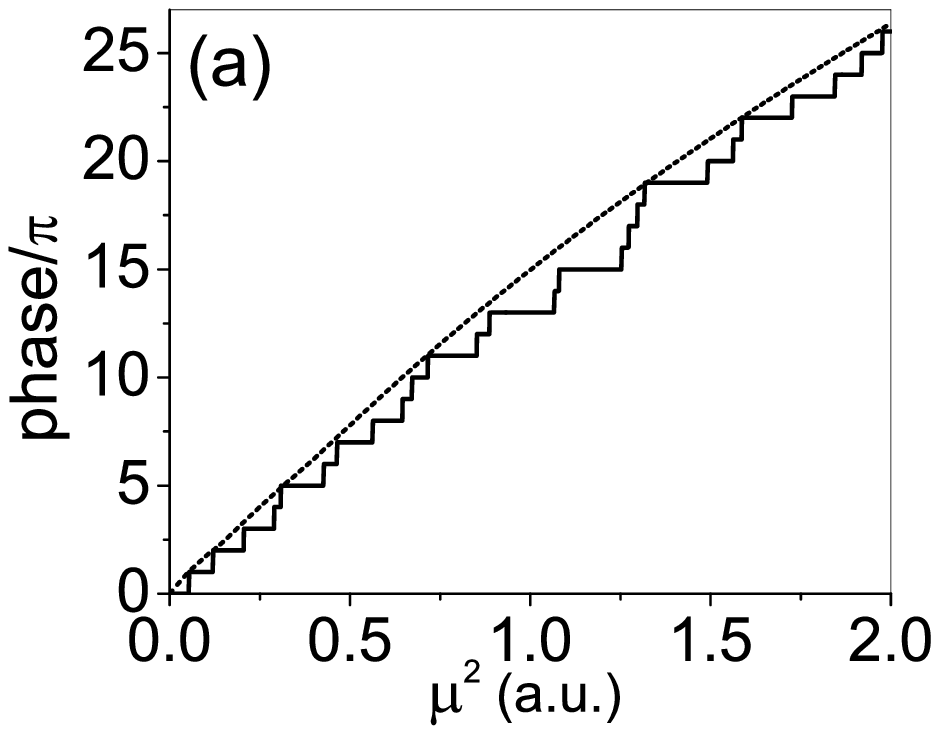}}
\centerline{\epsfxsize=7.0cm\epsfysize=7.0cm\epsfbox{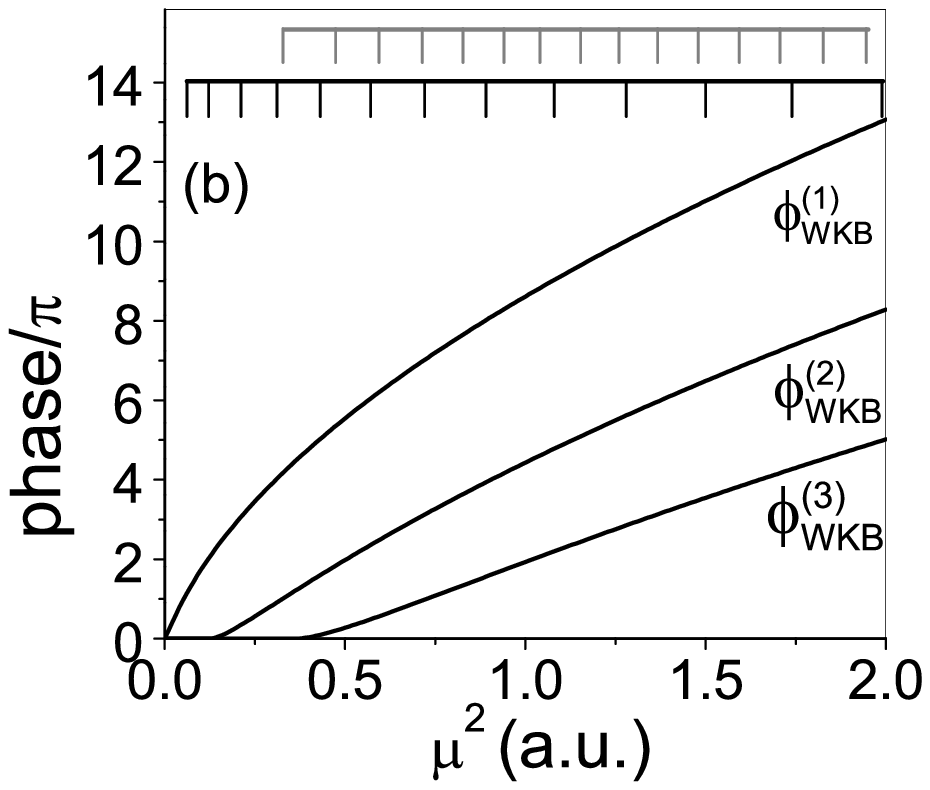}}
\caption{(a) Eigenphase (solid) and total adiabatic WKB phase (AWP,dashed) 
for the 
dipole-scattering model.  (b) AWP contributions for the lowest adiabatic 
curves ($\phi^{(1)}_{WKB}$) with the black bracket indicating when 
$\phi^{(1)}_{WKB}$ passes through
a multiple of $\pi$.  The two remaining curves are contributions with 
non-zero partial waves ($\phi^{(2,3)}_{WKB}$).  The gray bracket indicates 
where the sum of these contributions pass through a multiple of $\pi$.} 
\label{model}\end{figure}  
To show that the broad resonances primarily belong to the lowest
potential and the narrow shape resonances belong to the higher-lying
potentials, we use an eigenphase analysis.
The eigenphase can be thought of as the sum of the phase shifts for 
all of the channels; thus it tracks the behavior of all the 
channels simultaneously.  The eigenphase is defined as 
\begin{equation}
\phi_{eigen}=\sum_i \tan^{-1}(\lambda_i^{K}).
\label{eigen}
\end{equation}
Here $\lambda_i^{K}$ are the eigenvalues of the $K$ matrix from the
full-scattering calculation \cite{Taylor}.  (The $K$ matrix is
related to the more familiar scattering matrix by 
$S=(1+iK)/(1-iK)$.)  When the system gains 
a bound state, it appears as a $\pi$ jump in eigenphase.
The eigenphase of the system is shown in Fig. \ref{model} (a), as
the solid line with many abrupt steps.

To analyze this situation further, we construct an approximate
eigenphase as follows.  First, we asses the total phase accumulated
in each adiabatic channel, using a WKB prescription:
\begin{equation}
\phi_{WKB}^{(i)}(\mu)
=\int_{R_{in}}^{R_{out}}\sqrt{-2 m_r V_{AD}^{(i)}(\mu,R)/\hbar^2}.
\nonumber
\end{equation}
Here $(i)$ stands for the $i^{th}$ adiabatic curve.  For the lowest
adiabatic curve, which is always attractive, the range of integration
is from $R_{in}$ to $R_{out}=\infty$.  For high-lying channels that
possess a barrier to scattering at zero collision energy, the limits
of integration are from $R_{in}$ to the inner classical turning point
of the barrier.  This will yield some information on shape resonances
trapped behind the barrier, but we will not make much of this in
the analysis to follow.  Finally, we add together the individual
WKB phases to produce an approximate eigenphase shift, which we
dub the ``adiabatic WKB phase'' (AWP):
\begin{equation}
\phi_{WKB}(\mu)=\sum_i\phi_{WKB}^{(i)}(\mu),
\label{wkbphase}
\end{equation}
Since we are not concerned with properties of the phase associated 
with the higher-lying adiabatic curves, we do not consider the connection 
formula now.

The total AWP for this system is shown as a dashed line in 
Fig. \ref{model} (a).
It tracks the eigenphase well but offers more information if we
decompose the AWP into its contributions.
Figure \ref{model} (b) shows the individual contributions of the sum.
The largest contribution is the phase from the lowest adiabatic curve,
which can be associated with potential resonances.  A black bracket appears 
above this phase contribution with vertical marks indicating when 
it passes through an integer multiple of $\pi$, i.e., when we expect to 
see a potential resonance in the 
cross section.  This same bracket is plotted in Fig. \ref{model-sigma} and
shows good agreement between the locations of the potential resonances and 
the AWP predictions. We conclude from this that the main resonance features 
in the Stark spectrum arise primarily from this single potential curve.  

In Fig. \ref{model} (b) the two remaining phase contributions originate
from non-zero partial wave channels that possess centrifugal barriers;
see Fig. \ref{model-sigma}(b).  The gray 
bracket indicates where the sum of these two contributions pass through
an integer multiple of $\pi$, and thus represent a guess for where
the shape resonances lie.  This gray bracket is also shown in 
Fig. \ref{model-sigma}.  The agreement with the position of the
narrow resonance features in the cross section is not 
nearly as good as it is with the broad potential resonances.  
This indicates a more  involved
criterion for shape resonances.  Nevertheless, the AWP predicts 15 shape 
resonances, and
there are 13 in the range of $\mu$ shown.  The AWP appears to offer a means to
roughly predict the number of shape resonances in this system, even though 
it does not predict the locations exactly.  

A main point of this analysis is that the AWP in the lowest adiabatic
channel alone is sufficient to locate the potential (as opposed to shape)
resonances, without further modification.  For the rest of this paper, 
we focus on the potential resonances in more realistic molecules
with internal molecular structure.  The general analysis in terms of a 
single-channel
AWP will still hold, but an additional phase shift will be required to
describe the spectrum.

\section{Strong Field Seekers}
Strong-field-seeking molecules are approximately described as polarized
in the sense of the last section, because their dipole moments are
aligned with the field.   They will, however, contain a richer
resonance structure owing to the presence of low-lying excited states
that can alter the dipolar potential energy surface at small-$R$.

For concreteness, we focus here on molecules with a  $^1\Sigma$ ground state.
Heteronulcear alkalis fit into this category and are rapidly approaching
ground state production with various species \cite{alkali}.  
As examples, we pick RbCs and SrO in their ground states. Ground state
RbCs has been 
produced experimentally \cite{Sage}.  As for SrO, promising new
techniques should lead to experimental results soon \cite{Demille}.
For simplicity we include only the $J=0$, $J=1$ rotational states, 
and freeze the projection of molecular angular momentum to $M_J=0$.  
This restricts the number of scattering thresholds 
to three, identified by the parity quantum number of the molecules 
in zero field.  The parity quantum numbers for the three thresholds 
are ($--,-+,++$). This model is similar to the one presented in Ref. 
\cite{AA_PRA} which can be easily constructed for any rigid rotor when only 
including two molecular states. 
One immediate consequence of multiple thresholds is the presence 
of rotational Fano-Feshbach resonances in the collisional spectrum.  

\begin{figure}\centerline{\epsfxsize=7.0cm\epsfysize=7.0cm\epsfbox{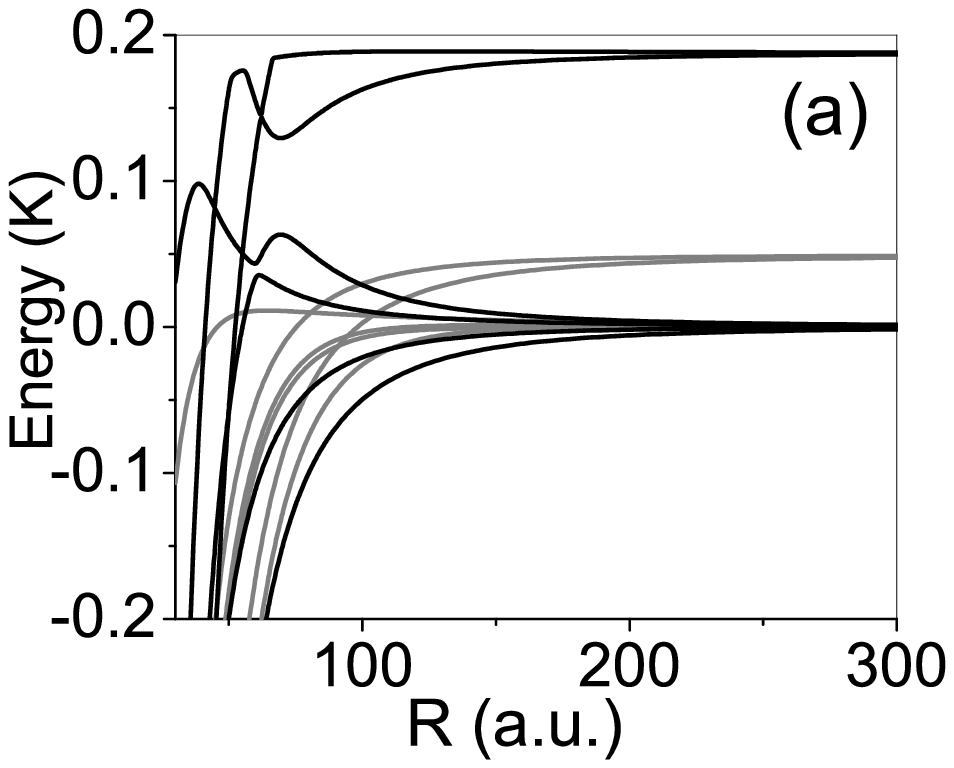}}
\centerline{\epsfxsize=7.0cm\epsfysize=7.0cm\epsfbox{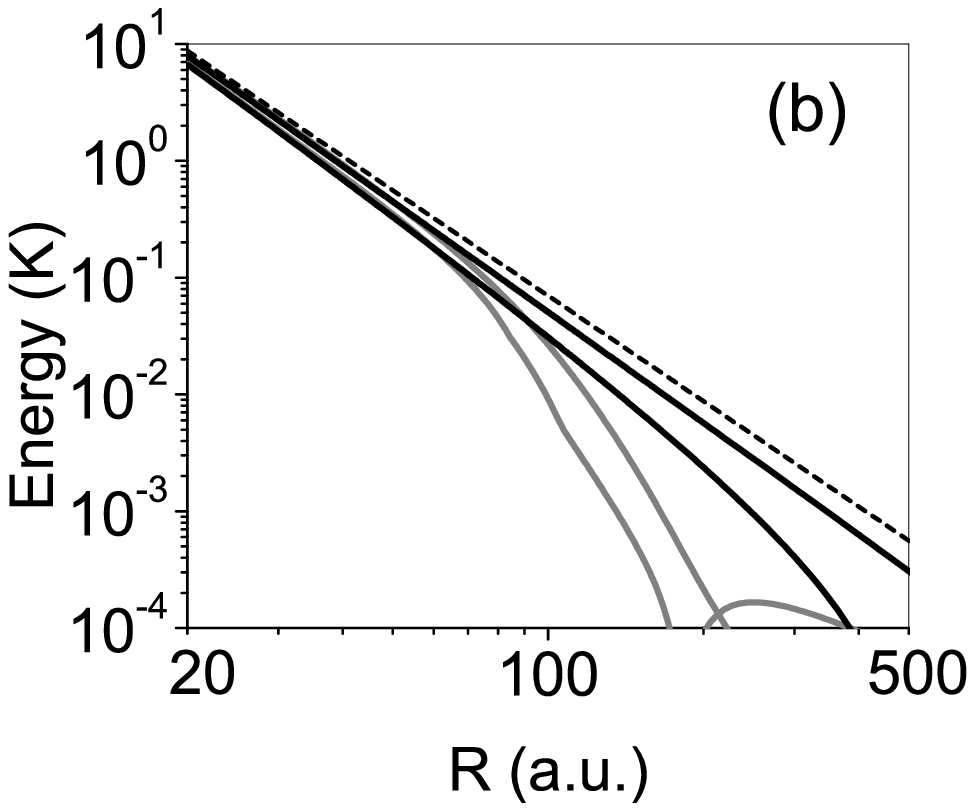}}
\caption{(a) Adiabatic curves for the RbCs system at two different
field values, ${\cal E}=0 (V/cm)$ (gray) and  ${\cal E}=5000 (V/cm)$ (black).
(b) Log-log plot of the absolute value of the two lowest adiabatic curves 
The blue dashed line is proportional to $1/R^3$.}\label{rbcs-ad}\end{figure}

The first example is RbCs, whose physical parameters are 
$\mu=1.3$ D, $m_r=110$ a.m.u.,  and $B_e=0.0245 (K)$ \cite{Sage}. 
As before, we apply vanishing boundary condition at $R_{in}=20$ a.u.
To converge this calculation over the field range considered, 
we require partial waves up to $l=30$.
We first look at the adiabatic curves of the system to get an understanding
of how the real system deviates from the simple model presented above.  
In Fig. \ref{rbcs-ad} (a) we plot the 6 lowest adiabatic curves for the
RbCs system with only four partial waves, so the figure is more easily 
interpreted.  The sets of adiabatic curves 
shown are for two different field values: the gray set has $\cal E$=0 and 
the black set has $\cal E$= 5000 (V/cm).  There are two important features
that differ from the dipole example. 
First, there are two higher thresholds, and the electric field shifts these 
apart in energy as the field is increased.  Second,the electric field 
dramatically changes the radial dependence of the Hamiltonian. 

The difference in thresholds can be seen clearly in Fig. \ref{rbcs-ad} (a)
where the lowest excited threshold moves from $0.05(K)$ 
at $\cal E$=0 to $0.18 (K)$ at $\cal E$=5000 (V/cm).
The difference in radial dependences for the two cases is seen more clearly 
in a log-log plot of the two lowest adiabatic curves for both fields,
as shown in Fig. \ref{rbcs-ad} (b).  The gray set corresponding to 
zero field has two distinct asymptotic radial powers.  At large $R$ the 
lowest adiabatic curve has a $1/R^6$ behavior asymptotically because of 
couplings with channels far away in energy [$\sim0.05 (K)$].  However as
$R$ approaches zero the dipolar interaction has overwhelmed the rotational
energy separation and the radial dependence becomes $1/R^3$ in character at 
about R=100 (a.u.). For reference, the dashed line is proportional to $1/R^3$.
With $\cal E$= 5000 (V/cm), the two black curves show the
radial dependence of the adiabatic curves.  The lowest curve now has nearly
$1/R^3$ over the whole range shown.  Asymptotically when the centrifugal 
barrier is larger, the radial dependence will change to $1/R^4$ 
\cite{AA_PRA}.  
The second lowest adiabatic curve is also significantly altered by the strong 
dipolar interaction as can be seen in Fig. \ref{rbcs-ad} (b).

\begin{figure}\centerline{\epsfxsize=7.0cm\epsfysize=7.0cm\epsfbox{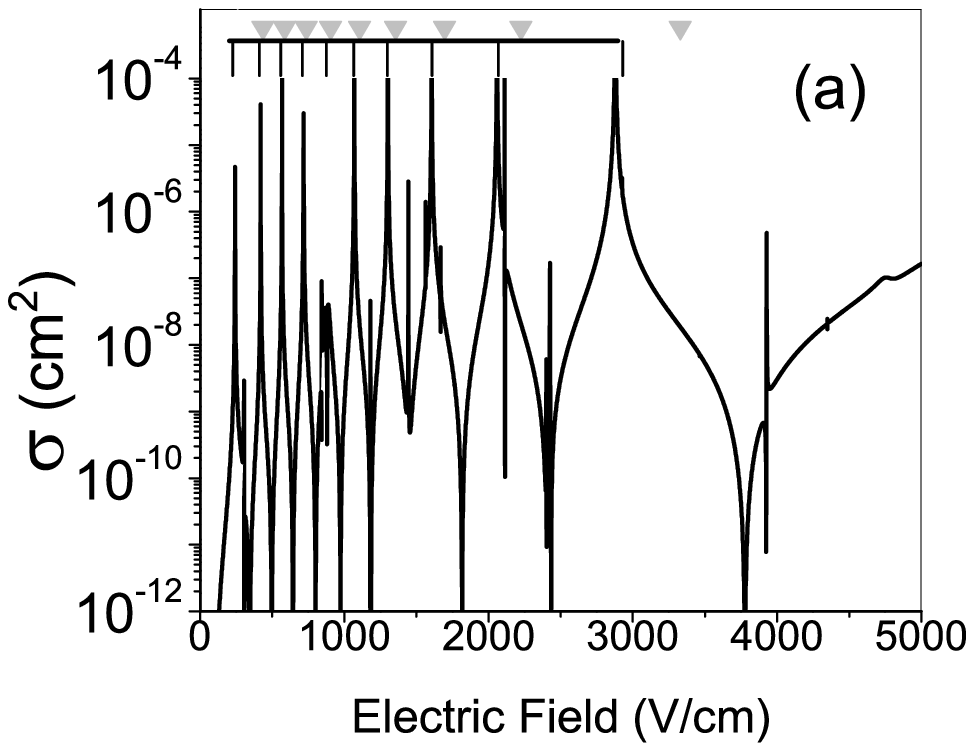}}
\centerline{\epsfxsize=7.0cm\epsfysize=7.0cm\epsfbox{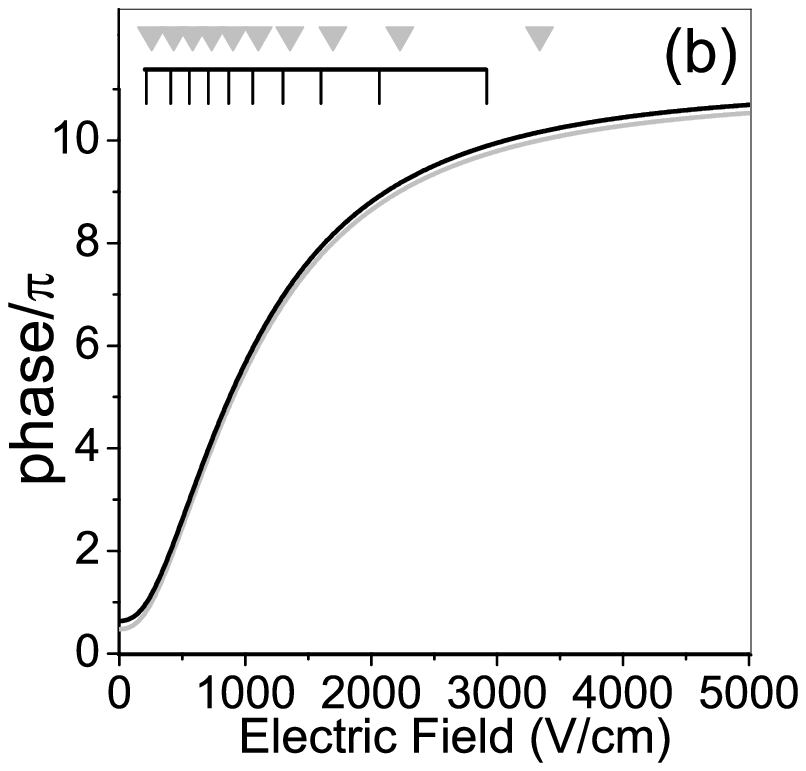}}
\caption{(a) Cross section for RbCs in strong-field seeking states,
including the 10 potential resonances that the AWP predicts (solid bracket).  
The AWP phases are shown in (b).  The black bracket corresponds to the AWP 
predicted resonances for the lowest adiabatic curve after a defect, 
$\delta_{defect}=0.14$, has been added according to Eq. (\ref{defect}).
The gray triangles are the AWP predicted resonances without the defect.
} \label{RB}\end{figure}

In Fig. \ref{RB} (a), we plot the cross section
for the model RbCs system.  This spectrum is riddled with narrow 
Fano-Feshbach resonances, but is still dominated by a series of
potential resonances
similar to the one in Fig. \ref{model-sigma} (b). There are two sets 
of AWP predictions shown as over-brackets.  To understand their 
difference, we look to Fig. \ref{RB} (b).  
The AWP for the lowest adiabatic curve is shown in Fig. \ref{RB} (b) for
two different zero-field phase values.  The gray curve is the AWP that is 
directly computed from the method described above, Eq. (\ref{wkbphase}).  
The locations where it 
passes through an integer multiple of $\pi$ are indicated by the gray 
triangles.  Referring back to Fig. \ref{RB} (a), where the same gray triangles
appear we see that this simple
estimate does {\it not} reproduce the resonance position.

We can, however, introduce an additional overall phase shift to
account for the difference in short-range interactions from the
pure polarized case.  The shifted AWP reads
\begin{equation}
\tilde\phi_{WKB}^{(1)}({\cal E})=\phi_{WKB}^{(1)}({\cal E})+\pi\delta_{defect}
\label{defect}.
\end{equation}
By treating $\delta_{defect}$ as a fitting parameter, we can obtain
the resonance positions indicated by the black bracket in Fig. \ref{RB},
which agree quite well with the resonance positions in the close-coupled
calculation.  To do so requires, in this case, a phase shift 
$\delta_{defect}=0.14$.
In analogy with Rydberg spectroscopy we consider the shift we have
added to be a  ``quantum defect'' that accounts for the effect of the 
short-range interaction.  The additional phase shift reflects the influence
of short range physics on the scattering, such as curve crossings with 
curves from higher thresholds.

The AWP also saturates with field, as can be seen in Fig. \ref{RB} (b).
This occurs because the electric field eventually fully polarizes 
the molecules, so the dipole moment cannot increase further.
The effect can also be seen in the spacing of the potential resonances.  
At low fields, the potential resonances occur frequently in field. 
Then, as the field is further increased, the
resonances occur less often in field, which is a signature of dipole moment
saturating and therefore an increasing field having less effect on the 
molecular interaction.

\begin{figure}\centerline{\epsfxsize=9.0cm\epsfysize=6.0cm\epsfbox{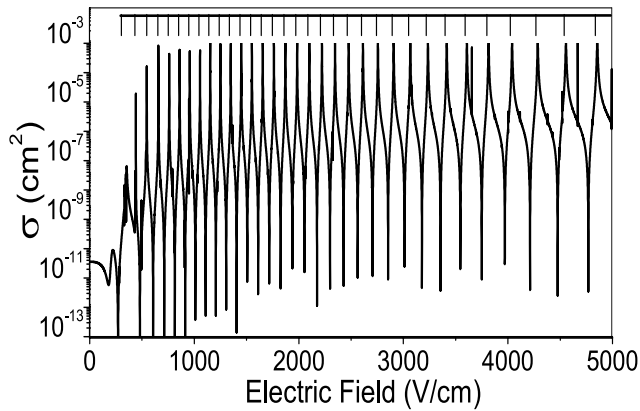}}
\caption{Cross section for SrO with a bracket indicating resonance 
positions predicted by the phase shifted AWP 
for the lowest adiabatic curve.  There are 33 PR shown in this range
of electric field. $\delta_{defect}=0.215$}
\label{sro}\end{figure}

As a second example, we consider SrO, which has the physical 
parameters  $m_r$=52 a.m.u, $\mu$=8.9 D, and $B_e=$ 0.5 K \cite{Demille}. 
We choose this molecule for its comparatively large mass and dipole
moment, which guarantee a large number of resonances.
Fig. \ref{sro} we have plotted the cross section for SrO, which is 
dominated by the quasi-regular potential resonance series. 
As before, the black bracket indicates where the phase shifted 
AWP predicts the potential resonances, and we see the agreement is
good.  Furthermore, the series has not terminated since we have not 
completely polarized the dipole.  The series of potential resonances 
saturates at 
17.5 (kV/cm) after a total of 43 potential resonances have been induced.
To line up the AWP's predictions and the actual cross section requires a 
defect of $\delta_{defect}=0.215$.

We have picked two examples to illustrate how the
potential resonances will appear in the context of collisional spectroscopy.
These resonance will occur to varying degrees in the strong-field seeking 
collisions of all polar molecules. For example, we can also make similar
predictions for 
an asymmetric rotor molecule such as formaldehyde $(H_2CO)$.
We find that this molecule should possess six potential 
resonances in the field interval from 0 to 50 $(kV/cm)$.

It is worth noting that portions of similar resonance series 
was anticipated in
cold atomic gases subjected to electric fields \cite{Deb}.  However very few
such resonances are likely to be observed, owning to the enormous fields
($\sim MV/cm$) required to generate them.  In polar molecules, by contrast, 
the entire series should be readily observable.

\section{Collisional Spectroscopy}
Through the course of this work we have shown that the zero-energy
cross section of weak-field-seeking molecules
is dominated by a set of broad potential resonances.  Even though these 
resonances are themselves intriguing, their properties can be exploited
to learn much more about the system.  The general structure of the potential 
resonances is governed by the long range dipolar interaction, which has a 
predictable and common form.  With a clear understanding of this interaction
and how it induces resonances, it could be exploited to learn about the 
short range interaction of the molecules. This is because
details of where the lines appear must also
depend on the boundary condition experienced by the wave function at small
values of $R$.  Therefore the spectrum contains information on the 
small $R$ intermolecular dynamics.  Thus by studying the potential resonance
series we can extract information about the short range dynamics.  

This idea is similar to quantum defect theory, which, for example,
has been very successful in Rydberg spectroscopy.  The short range physics 
of the electron interacting with the nucleus is complicated 
and not easily solved.  However
once the electron is out of the small $R$ region, it enters into a pure
coulomb potential where the motion of the electron is well understood.
The effect of the short range must be merged with the long range physics to 
form a complete solution.  To account for the short range interaction, the 
energy can be parameterized by replacing the principal quantum
number by an effective quantum number $n^\star=n-\mu$.  This
procedure is tantamount to identifying an additional phase shift
due to the interaction of the electron with the atomic core.
The idea of merging standard long range physics with complicated short
range behavior has been applied successfully not only in Rydberg states of 
atoms \cite{qdt,Fano} and molecules \cite{Jungen}, but also in atomic 
collisions \cite{Mies}, cold collisions \cite{Greene2,Mies2,Gao}, 
and dipole-dipole interactions of the type we envision here \cite{Deb}.

\begin{figure}\centerline{\epsfxsize=9.0cm\epsfysize=8.0cm\epsfbox{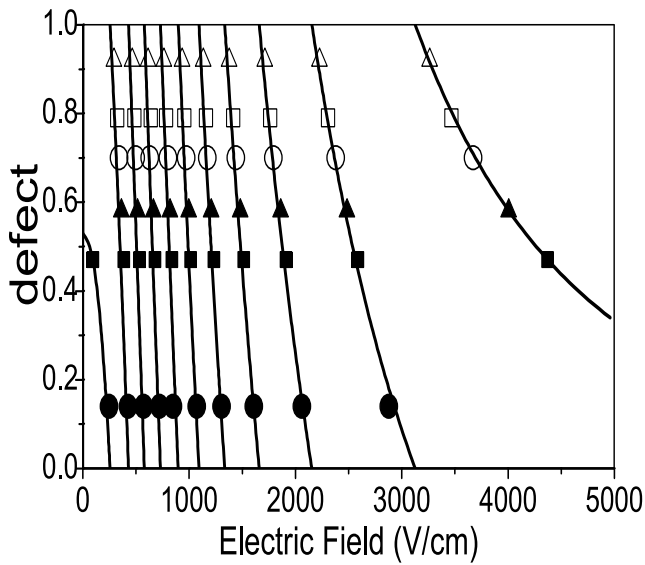}}
\caption{The vertical curves represent electric field values where the AWP 
predicts potential resonances for a given defect,
${\cal E}_{WKB}(\delta_{defect})$.
The points represent resonant field values, ${\cal E}^{(b)}$, in the 
full calculations with different initial boundary conditions.
The values of $b$ are given by 
cotan$(\pi\beta)$ and $\beta$ for the different full numerical
calculations are $0$ (filled circle),
$0.11$ (filled square), $0.22$ (filled triangle), $0.56$ (hollow circle),
$0.78$ (hollow square), and $0.89$ (hollow triangle).
Each set of values  ${\cal E}^{(b)}$ is plotted at a height corresponding to 
its best-fit value of $\delta_{defect}$.}\label{defect-fig}\end{figure}

As a simple expression of this idea, we can alter the boundary
condition applied at $R_{in}$ when performing the full scattering
calculation and note its influence on the field dependent spectrum.  
For the above calculations we have imposed the standard vanishing boundary 
condition, $\psi(R_{in})=0$ for all channels.
We now replace this condition with a uniform logarithmic derivative,
$b=({d\over dR}\psi)\psi^{-1}$, at $R_{in}$.  Thus previously we set 
$b=\infty$, but now we allow $b$ to vary.  The log-derivative 
is conveniently represented as a phase:
\begin{equation}b={\rm cotan}(\pi\beta),\end{equation}
where $\beta$ can lie between zero and one, covering all values of 
$b$ from $-\infty$ to $+\infty$. For $\beta=0$, the boundary 
condition is the one employed above, $\psi(R_{in})=0$, whereas for
$\beta=0.5$, the boundary condition is ${d\over dR}\psi(R_{in})=0$. 

We have re-computed the collisional spectrum of RbCs for several 
different initial 
conditions, and plotted the field values of the potential resonances, 
${\cal E}^{(b)}$, in Fig. \ref{defect-fig} as sets of points. 
The values of $\beta$ for the different calculations are $0$ (filled circle),
$0.11$ (filled square), $0.22$ (filled triangle), $0.56$ (hollow circle),
$0.78$ (hollow square), and $0.89$ (hollow triangle).
The filled circles are resonant locations for cross section in 
Fig. \ref{RB} (a).

We next wish to demonstrate that each such spectrum can be 
identified by a single quantum defect parameter, as was done in
the previous section.  This entails picking a value
of $\delta_{defect}$ and then setting the  phase 
$\tilde\phi_{WKB}^{(1)}({\cal E})$ equal to
an integer multiple of $\pi$.  This yields a set of resonant field
values, ${\cal E}_{WKB}(\delta_{defect})$.  Each $\delta_{defect}$ corresponds 
to a particular approximate spectrum.  The set of curves,
${\cal E}_{WKB}(\delta_{defect})$, generated by continuously varying 
$\delta_{defect}$ are shown are shown in Fig. \ref{defect-fig} as
solid lines.  The bracket in  Fig. \ref{RB} (a)
corresponds to the set of points where a vertical line intersects 
${\cal E}_{WKB}(\delta_{defect})$ with $\delta_{defect}=0.14$.

We can compare the the resonant field values predicted by the AWP, 
${\cal E}_{WKB}(\delta_{defect})$, and resonant field values given by
the full calculation with different initial conditions, ${\cal E}^{(b)}$.  
To plot ${\cal E}^{(b)}$ we have varied the height at which the set of points
${\cal E}^{(b)}$ is plotted until it aligns with
${\cal E}_{WKB}(\delta_{defect})$.   Doing this we are able to 
to see how $b$ and $\delta_{defect}$ are related.  Thus 
in fig. \ref{defect-fig} we can see that even with different boundary 
conditions, the single AWP curve in fig \ref{RB} (b) can be used to predict 
the spacing between the potential resonances by varying a single parameter,
$\delta_{defect}$.  This shows 
the AWP represents the long range scattering physics well,
and that empirically extracted parameters like $\delta_{defect}$ will carry 
information about the short-range physics such as that embodied in $b$.

\section{Conclusion}
A number of resonant processes may occur when two polar molecules meet in 
an ultracold gas.  We have focused here on the dominant, quasi-regular,
series of potential resonances between weak-field seeking states.  
These potential resonances originate in the
direct deformation of the potential energy surface upon which the molecules 
scatter.  Observation of these resonances may offer a direct means for 
probing the short range interaction between molecules.  We have provided a 
means of analyzing this system with an adiabatic WKB phase integral.  This 
method shows how the system evolves with the application of an electric field.

\begin{acknowledgments}
This work was supported by the NSF and by a grant from the
W. M. Keck Foundation.  The authors thank D. Blume 
for a critical reading of this manuscript.
\end{acknowledgments}

\bibliographystyle{amsplain}

\end{document}